\documentclass[a4paper]{jpconf}
\usepackage{graphicx}
\begin{document}
\title{Neutrino induced charged-current coherent $\rho$ production}

\author{X. C. Tian}

\address{Department of Physics and Astronomy, University of South
  Carolina, Columbia, SC, 29208, U.S.A. }

\ead{tianx@mailbox.sc.edu}

\begin{abstract}
We present the latest results of coherent $\rho$ (Coh$\rho$) production using the large data set collected by the NOMAD detector in which the momenta, charges, and photons are precisely measured. We discuss the application of using Coh$\rho$ process to constrain the neutrino flux with the proposed Long-Baseline Neutrino Experiment Near Detector,  the high resolution Straw Tube Tracker. %The charged or neutral $\rho$ meson can be produced coherently in the neutrino-nucleus interactions. 
\end{abstract}

\section{Introduction}
With the discovery of non-zero $\theta_{13}$ by the reactor and accelerator neutrino experiments~\cite{dayabay, reno} the focus has been shifted to the determination of neutrino mass hierarchy and the measurement of the leptonic CP violation phase, which may hold the key to the matter-antimatter asymmetry in the universe. To correctly interpret the oscillation data one needs to understand the neutrino-nucleon (nucleus) interaction thoroughly. The low momentum ($Q^2$) transfer, high energy ($\nu$) transfer interactions of neutrino induced coherent $\rho$ meson production allows us to study the basic properties of the weak current: the Conservation of the Vector Current hypothesis (CVC)~\cite{feynman,adler}. 

%\begin{figure}[h]
%\begin{minipage}{14pc}
%\begin{center}
%\includegraphics[width=14pc]{figures/cohrhop}
%\caption{\label{label}Figure caption for first of two sided figures.}
%\end{minipage}\hspace{2pc}%
%\begin{minipage}{14pc}
%\includegraphics[width=14pc]{figures/cohrhom}
%\caption{\label{cohrho_diag} The (anti-)neutrino induced charged-current coherent $\rho$ production on nucleus.}
%\end{minipage} 
%\end{center}
%\end{figure}
 
%\Fref{cohrho_diag} shows the Feynman diagram of coherent $\rho^\pm$ process. 
Based on the vector-dominance model~\cite{piketty}, the differential cross section can be derived as~\cite{kopeliovich}:
  \begin{equation}
    \frac{d^3\sigma (\nu_\mu {\cal A} \rightarrow \mu^- \rho^+ {\cal A} )}{dQ^2 d\nu dt} =
    \frac{G_F^2}{4\pi^2}    \frac{f^2_\rho}{1-\epsilon}     \frac{|q|}{E_\nu^2} 
    \left [  \frac{Q}{Q^2 + m^2_\rho}  \right ]^2 
    (1 + \epsilon R) \left [ \frac{d \sigma^T (\rho^+ {\cal A} \rightarrow \rho^+ {\cal A})}{dt} \right ],
    \label{eq-cohrhop2}
  \end{equation}
  
  \noindent
  where $G_F$ is the weak coupling constant,
  $Q^2 = -q^2 = -(k-k')^2$, $t =  (p-p')^2$, $\nu = E_\nu - E_\mu$, $x = Q^2/2M\nu$,
  $y=\nu / E_\nu$.
  $f_\rho$ is related to the $\rho$ form-factor,
  the polarization parameter
  $\epsilon = \frac{4 E_\nu E_\mu - Q^2}{4E_\nu E_\mu + Q^2 + 2 \nu^2}$,
  and $R=\frac {d\sigma^L/dt}{d\sigma^T/dt}$ with $\sigma^L$ and $\sigma^T$ as
  the longitudinal and transverse $\rho$-nucleus  cross sections.
  The $\rho$ form factor, $f_\rho$, is related to the
  corresponding factor in charged-lepton scattering,
  $f_{\rho^{\pm}}  = f^\gamma_{\rho^0} \sqrt{2} \cos\theta_C$, where $\theta_C$ is the Cabibbo angle and
  $f^\gamma_{\rho^0} = m^2_{\rho}/\gamma_\rho$
  is the coupling of $\rho^0$ to photon ($\gamma^2_\rho/4\pi = 2.4 \pm 0.1$).
  
  \noindent
  Following the Rein-Sehgal model of meson-nucleus absorption~\cite{rein},
  \begin{equation}
    \frac{d \sigma^T (\rho^+ {\cal A} \rightarrow \rho^+ {\cal A})}{dt}  =
    \frac { {\cal A}^2 }{16 \pi}  \sigma^2({\rm h n}) \exp(-b |t|) F_{\rm abs},
    \label{eq-rs-abs}
  \end{equation}
  
  \noindent
  where $\sigma (\rm h n)$ is the ``hadron-nucleon'' cross-section with the
  energy of the hadron $\simeq \nu$, $b = R^2/3$ such that $R=R_0 {\cal A}^{1/3}$,
  with $R_0 = 1.12 \,\rm fm$ and the absorption factor $F_{\rm abs} = 0.47 \pm 0.03$.

\section{The NOMAD Experiment}
The NOMAD experiment is a short baseline neutrino experiment designed to search for $\nu_\mu\rightarrow \nu_\tau$ oscillations using the
CERN Super Proton Synchrotron (SPS) wideband neutrino beam~\cite{nomad}. The 450 GeV protons from the SPS impinge on a beryllium target producing secondary mesons ($K$, $\pi$ and $K_L^0$) which will produce neutrinos. The neutrino events dominated by $\nu_\mu$ ($\nu_\mu:\bar{\nu}_\mu:\nu_e:\bar{\nu}_e$=$1.00:0.025:0.015:0.0015$) have a mean energy  of $\sim$24 GeV spanning the region of  ${\cal{O}}(1)\leq E_\nu\leq 300$ GeV. The NOMAD detector consists of several sub-detectors.  132 planes of $3 \times 3$~m$^2$ drift chambers (DC) with an 
average density similar to that of  liquid hydrogen (0.1~g/cm$^3$) serve as the active target. The fiducial mass of 
the DC is 2.7 tons with an effective atomic number, $A=12.8$, similar to carbon. The DC provides excellent momentum
resolution, $\sigma_p/p = 0.05/\sqrt{L({\rm m})} + 0.008p({\rm GeV})/\sqrt{L(\rm m)^5}$. The DC is followed by 
nine modules of transition radiation detectors (TRD), a pre-shower (PRS) and a lead-glass electromagnetic calorimeter 
(ECAL). The TRD, PRS, and ECAL sub-detectors provide a high-resolution efficiency and purity  ($\geq 90$\%) of electron 
detection.  A dipole magnet provides a 0.4 T magnetic field orthogonal to the neutrino beam direction that surrounds the DC, 
TRD, and PRS/ECAL. The magnet is followed by hadron calorimeter (HCAL) and two muon-stations comprising large area 
drift chambers separated by an iron filter placed at 8- and 13-$\lambda$'s downstream of the ECAL, and the muon-stations provide a clean 
identification of the muons.  

The data used in this analysis were collected from 1995 to 1998 with $5.1\times10^6$ protons on target in neutrino mode, which corresponds to $1.4\times10^6$ 
$\nu_\mu$-CC interactions in the drift chamber. The Monte Carlo simulation of the coherent diffractive $\rho$ process
was done based on the diffraction model~\cite{piketty, kopeliovich,hyett}. The details of data and MC have been described elsewhere~\cite{qun}.

\section{Event Selection and Reconstruction}
The $\nu_\mu+\cal{N}\rightarrow \mu^-\cal{N}\rho^+$ followed by $\rho^+\rightarrow \pi^+\pi^0$ has two tracks ($\mu^-$ and $\pi^+$) originating from the reconstructed primary vertex. The $\pi^0$ promptly decays into two photons. The two photons either both convert in the drift chamber (2 $V_0$s event), or one of the photons converts in the drift chamber and the other is measured in the electromagnetic calorimeter (1 $V_0$ 1 cluster event), or both photons are measured in the ECAL (2 cluster event). 

The two-track Monte Carlo events with an identified muon within the fiducial volume that pass the following preselection cuts are used to build a likelihood ratio to separate the signal from background.
These preselection cuts are 1) the opening angle between $\mu^-$ and $\pi^+$ ($\theta_{\mu^-,\pi^+}$) is less than 0.5 rad, 2) the $\pi^+$ energy ($E_{\pi^+}$) is greater than 1 GeV, 
3) the missing transverse momentum ($p_{\rm m}^{\rm T}$) is less than 0.5 GeV, and 4) the event should have either 2 $V_0$s, 1 $V_0$ plus 1 cluster, or 2 clusters.

\section{The Results}
\begin{figure}[h]
\begin{center}
\begin{minipage}{16pc}
\includegraphics[width=16pc]{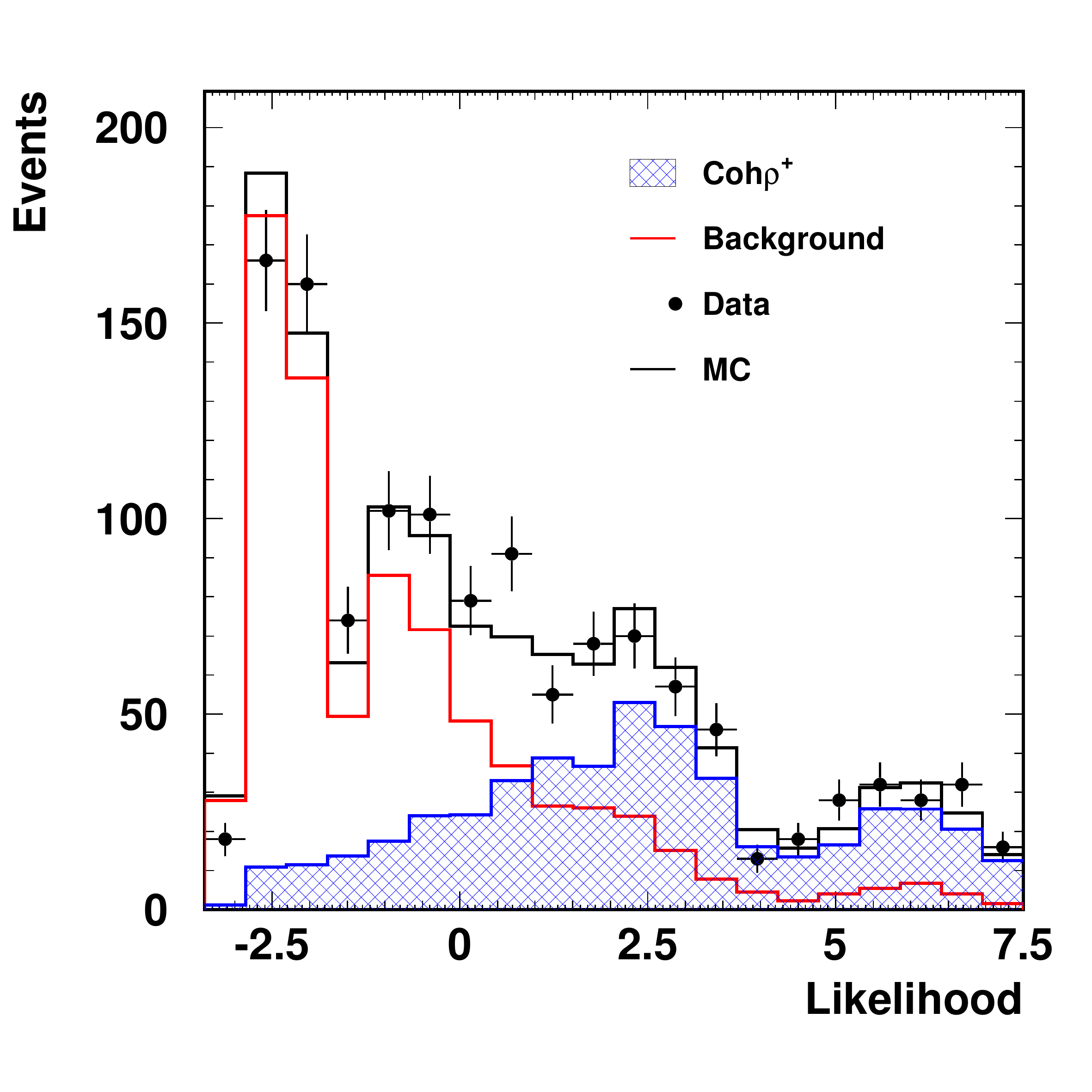}
\caption{\label{lh}The likelihood $\cal{L}$ distribution for data (points with error bars) and MC.}
\end{minipage}\hspace{2pc}%
\begin{minipage}{16pc}
\includegraphics[width=16pc]{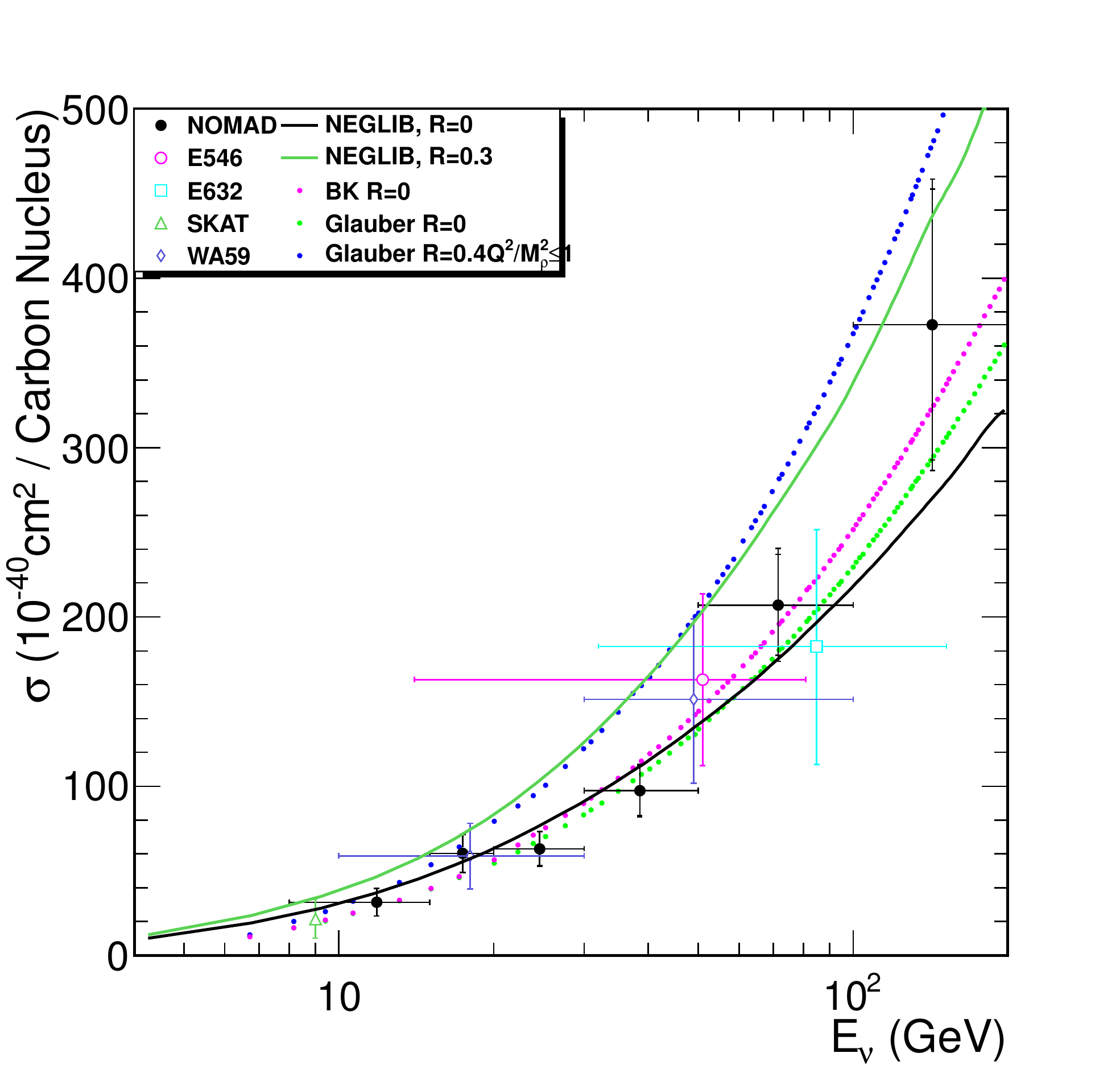}
\caption{\label{fig_summ} The Coh$\rho^+$ cross section as a function of incoming neutrino energy with different measurements and predictions.}
\end{minipage}
\end{center} 
\end{figure}

 The Bjorken variables $x$, $y$ and $\mathcal{Z}_{\pi^+}$ defined as $E_{\pi^+}\times(1-\cos\theta_{\pi^+})$ are used to construct the likelihood ${\cal{L}}=\ln\frac{P(l\vert\rm Coh\rho^+)}{P(l|\rm BKG)}$ to select the signal events. \Fref{lh} shows the likelihood distribution, and the signal region is defined as ${\cal{L}}>1.8$. An independent analysis using the neural network (NN) under the ROOT framework~\cite{root} gives very consistent results. \Tref{results} shows the numbers (background, raw data, efficiency and fully corrected Coh$\rho^+$ signal events) in the signal region obtained from both likelihood and NN analysis.  The total systematic uncertainty is $\pm$3.9\% which comes from 1) $\pm$1.6\% background normalization error dominated by charged-current Deep Inelastic events which is obtained by using the control region in likelihood or NN, 2) $\pm$2.5\% absolute normalization error which is from the inclusive charged current cross section measurement~\cite{qun}, and 3) $\pm$2.5\% signal efficiency related error which is found by varying the likelihood or NN cut, difference between likelihood and NN, and mock data study.

Using the measured inclusive $\nu_\mu$-CC cross-section from~\cite{qun} as a function of $E_\nu$, the absolute cross-section of Coh$\rho^+$ production
for A = 12.8 at the average energy of the neutrino flux $E_\nu=24.8$ GeV is determined to be:
\begin{equation}
\sigma(\nu_\mu+{\cal{N}}\rightarrow \mu^-{\cal{N}}\rho^+)=[67.1\pm4.8\,(\rm stat)\pm2.6\,(\rm syst)]\times10^{-40} \,\rm cm^2/nucleus,
\end{equation}
which is consistent with the prediction and previous measurements but with much better precision. \Tref{summary} is a compilation of the measurements up to today. \Fref{fig_summ}
shows the Coh$\rho^+$ cross section as a function of incoming neutrino energy with other measurements and predictions based on different models~\cite{digitize}. NOMAD data agree with the CVC and VDM based model prediction fairly well. The measurement favors the model with $R=0$, {\it i.e.} there is little longitudinal contribution in Coh$\rho^+$ production.

%\begin{figure}[h]
%\begin{center}
%\includegraphics[width=14pc]{figs/data_lh_xbj_p34}
%\includegraphics[width=14pc]{figs/data_lh_ybj_p34} \\
%\includegraphics[width=14pc]{figs/data_lh_zpi_p34}
%\includegraphics[width=14pc]{figs/data_lh_p35}
%\caption{\label{nn}The $x_{\rm bj}$, $y_{\rm bj}$, $p_{\rm m}^{\rm T}$ and  likelihood distributions for data (points with error bars) and MC.}
%\end{center}
%\end{figure}

\begin{table}[h]
\caption{\label{results} The results are obtained based on two independent analysis algorithms.}
\begin{center}
\lineup
\begin{tabular}{l|lll|l}
\br
        Algorithm & Background & Data & Efficiency & Coh$\rho^+$ Signal \cr
        \mr
        Likelihood & 86.1 & 363 & 0.064 & 4318.8$\pm$307.4 \cr \mr
        NN         & 76.1 & 356 & 0.065 & 4332.0$\pm$319.4 \cr
\br
\end{tabular}
\end{center}
\end{table}

\begin{table}[h]
\small{
\caption{\label{summary}A compilation of the charged-current coherent $\rho$ measurements.}
\begin{center}
\lineup
\begin{tabular}{l|llll|l}
\br
Experiment & $\nu/\bar{\nu}$ & Channel    & Target        & $<E_\nu>$ (GeV) & $\sigma$ ($10^{-40}$ cm$^2$/nucleus)\cr
\mr
E546~\cite{e546}                 & $\nu$                   & $\rho^+$    & Neon  (A=20)      &   51             & 189.7$\pm$59 \cr
BEBC WA59~\cite{wa59nu,wa59nubar}   & $\bar\nu$           & $\rho^-$      & Neon (A=20)       &   18             & 73$\pm23$ \cr
E632~\cite{e632}                 & $\nu+\bar{\nu}$ & $\rho^\pm$ & Neon (A=20)       &   86            & 210$\pm$80 \cr
SKAT ~\cite{skat}                 & $\nu$           	  & $\rho^+$     & Freon (A=30)       &   10            & 29$\pm$16 \cr
NOMAD (This Exp.)& $\nu$            	  & $\rho^+$     & Carbon (A=12.8) & 24.8          & 67.1$\pm$5.4 \cr
\br
\end{tabular}
\end{center}
}
\end{table}

\begin{figure}[h]
\begin{center}
\includegraphics[width=14pc]{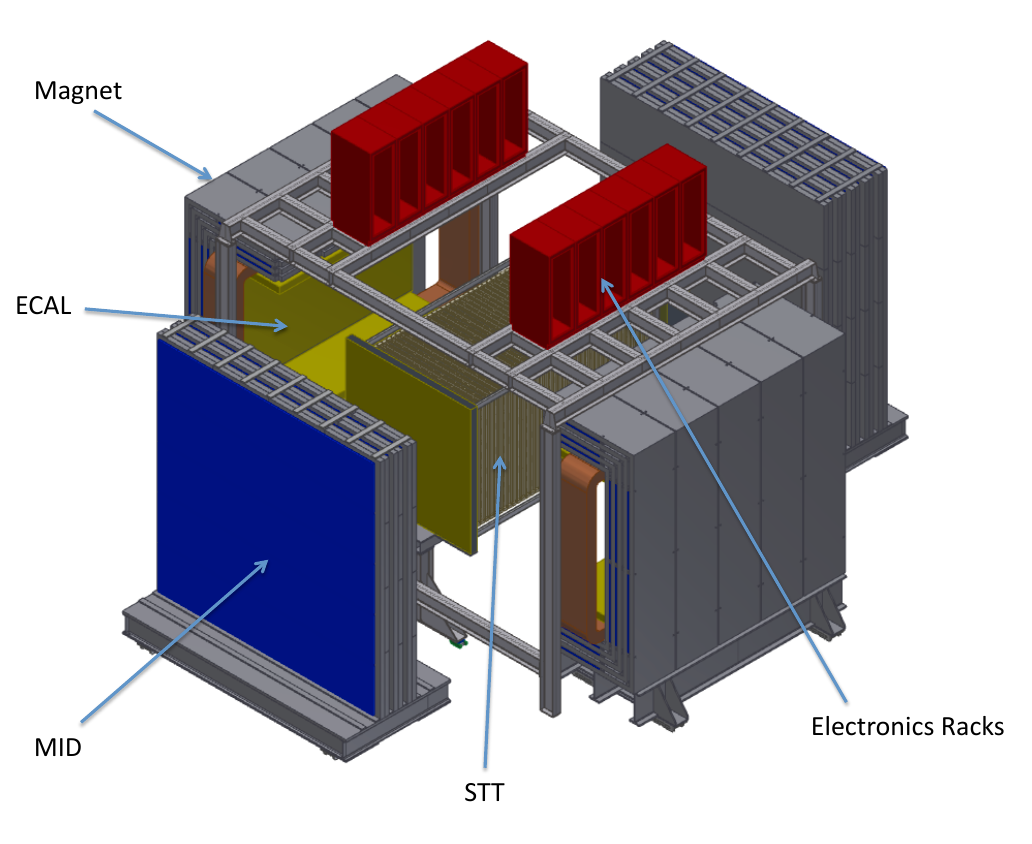}
\includegraphics[width=14pc]{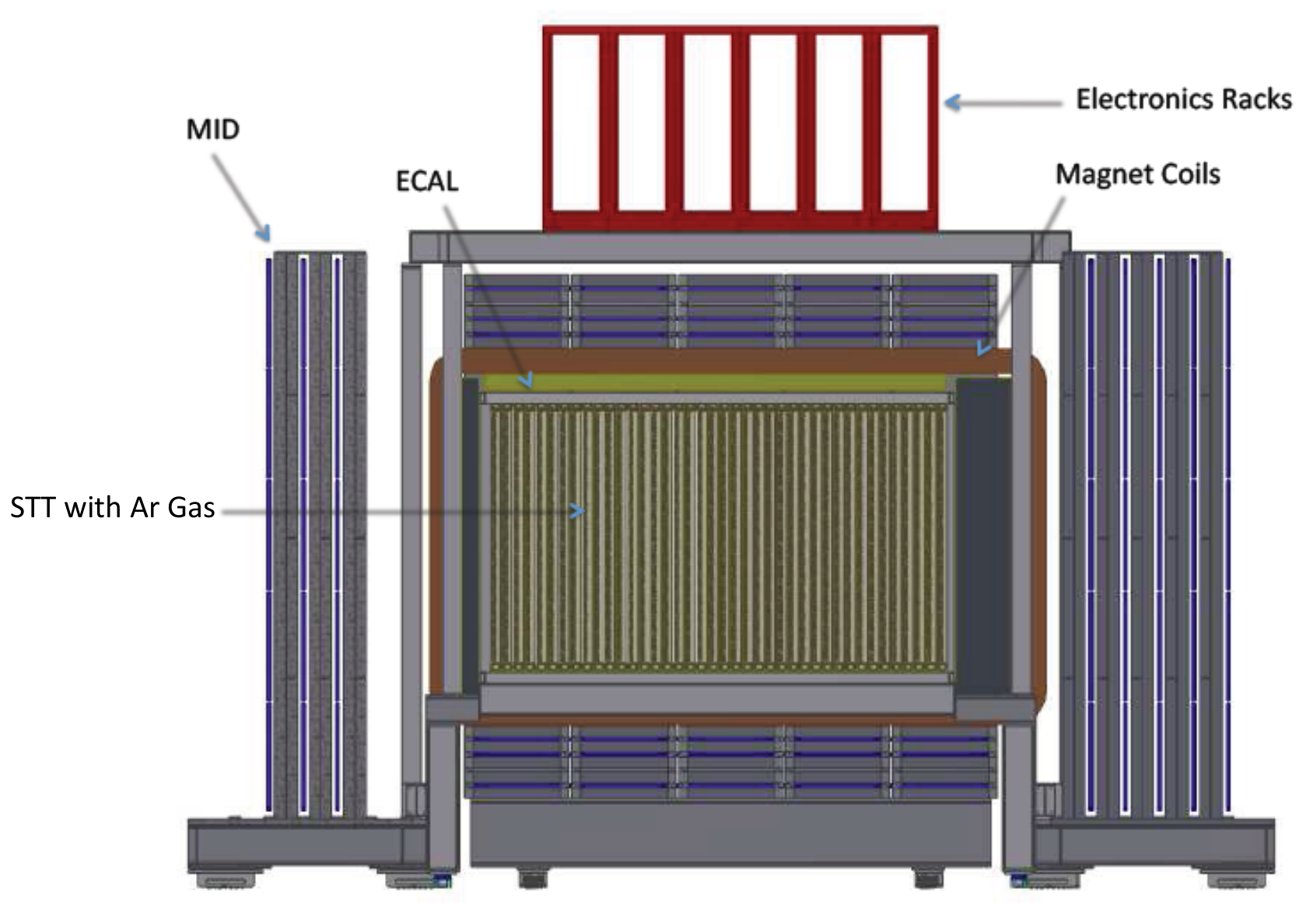}
\caption{\label{figs_hiresmnu}The proposed LBNE Near Detector - high resolution straw tube tracker.}
\end{center} 
\end{figure}

\vspace{-6mm}
\section{Discussion}
The precision determination of relative and absolute neutrino flux is essential to any neutrino oscillation and cross section measurements, especially in the precision era.
The primary goal of the Long-Baseline Neutrino Experiment (LBNE) is to determine the neutrino mass hierarchy and to observe CP violation.  The proposed high-resolution LBNE Near Detector~\cite{hiresmnu1, hiresmnu2} (\Fref{figs_hiresmnu}) could make measurements that would significantly  enhance LBNE's overall sensitivity. Measurements required to achieve LBNE's primary objectives include the relative abundance and energy spectra of all four neutrino species, $\nu_\mu$, $\bar{\nu}_\mu$, $\nu_e$, $\bar{\nu}_e$; neutrino and antineutrino cross-sections.
Besides the insight into the weak current by studying the Coh$\rho^+$ interactions, one can use this process to measure the absolute neutrino flux with high precision under the scope of CVC.
The LBNE Near Detector will have a much higher resolution and statistics, which permits us to better understand the space-time structure of the weak current, and to advance our understanding of the neutrino flux.


\begin{thebibliography}{9}
\bibitem{digitize} The Glauber and BK predictions are from the digitization of Ref.~\cite{e632}, then scaled to the NOMAD target
\end{thebibliography}
\end{document}